\documentclass[aps,pra,reprint]{revtex4-2}%
\usepackage{physics,verbatim,ulem,mathrsfs}
\usepackage{braket}
\usepackage{graphicx}
\usepackage{bm}
\usepackage{color}
\usepackage{subfigure}
\usepackage{ulem}
\usepackage{amsmath}
\usepackage{amssymb}
\usepackage{mathrsfs}
\usepackage{physics}
\usepackage[colorlinks,allcolors=blue]{hyperref}
\usepackage[toc,page,title,titletoc,header]{appendix}
\usepackage{dcolumn}
\usepackage{graphicx}
\usepackage{times}
\usepackage{natbib}
\usepackage{float}
\usepackage[english]{babel}
\usepackage{geometry}

\begin{document}

\title{Spin Squeezing with Arbitrary Quadratic Collective-Spin Interaction}
\author{Zhiyao Hu$^{1,2}$}
\thanks{These authors contributed equally to this work.}
\author{Qixian Li$^{1}$}
\thanks{These authors contributed equally to this work.}
\author{Xuanchen Zhang$^{1}$}
\author{Long-Gang Huang$^{1,3}$}
\author{He-bin Zhang$^{1}$}
\author{Yong-Chun Liu$^{1,4}$}
\email{ycliu@tsinghua.edu.cn}
\affiliation{$^{1}$
 State Key Laboratory of Low-Dimensional Quantum Physics, Department of
Physics, Tsinghua University, Beijing 100084, China
}%
\affiliation{
 $^{2}$School of Physics, Xi'an Jiaotong University, Xi'an 710049, China
}

\affiliation{$^{3}$China Fire and Rescue Institute, Beijing 102202, China}

\affiliation{$^{4}$Frontier Science Center for Quantum Information, Beijing
100084, China}
\date{\today }

\begin{abstract}
Spin squeezing is vitally important in quantum metrology and quantum
information science. The noise reduction resulting from spin squeezing can
surpass the standard quantum limit and even reach the Heisenberg Limit (HL)
in some special circumstances. However, systems that can reach the HL are
very limited. Here we study the spin squeezing in atomic systems with a
generic form of quadratic collective-spin interaction, which can be
described by the Lipkin-Meshkov-Glick(LMG) model. We find that the squeezing
properties are determined by the initial states and the anisotropic
parameters. Moreover, we propose a pulse rotation scheme to transform the
model into two-axis twisting model with Heisenberg-limited spin squeezing.
Our study paves the way for reaching HL in a broad variety of systems.
\end{abstract}

\maketitle

\sloppy
\preprint{APS/123-QED}




\section{Introduction}

Squeezed spin states (SSSs) \cite{1, 2} are entangled quantum states of a
collection of spins in which the uncertainty of one spin component
perpendicular to the mean spin direction is reduced below the standard
quantum limit (SQL). Owing to its property of reduced spin fluctuations, it
has a variety of applications in the study of many-body entanglement \cite%
{3,4,5,6,7,8,9,10,35,36}, high-precision measurements \cite{11,12,13,14,15,16,17},
and quantum information science \cite{18,19,20,21}. Many methods have been
proposed to realize spin squeezing, such as atom-light interaction \cite{22}%
, and quantum nondemolition measurement \cite{23}. One important way to
deterministically generate spin squeezing is utilizing the dynamical
evolution of squeezing interaction, which is accomplished via
collective-spin systems with nonlinear interaction \cite{2}. Typical
squeezing interactions include one-axis twisting (OAT) interaction and
two-axis twisting (TAT) interaction. The noise reduction of the TAT model
can reach the Heisenberg limit (HL), but the physical realization of the TAT
model is difficult. It is shown that the OAT model can be transformed into
TAT model using repeated Rabi pulses \cite{24}, but the more general cases
with other types of quadratic collective-spin interaction are still unknown.

Except for OAT and TAT interactions, the more general form of quadratic
collective-spin interaction can be described by the Lipkin-Meshkov-Glick
(LMG) model. The LMG model was first introduced in nuclear physics \cite%
{25,26,27,28,29,30}, which provides a simple description of the tunneling of
bosons between two degenerate levels and can thus be used to describe many
physical systems such as two-mode Bose-Einstein condensates \cite{31} or
Josephson junctions \cite{32}. A recent study shows that the LMG model could be used in generating spin squeezing and having 6.8(4) dB metrological gain beyond the Standard Quantum Limit with suitable time-reversal control in cavity QED \cite{33}. However, it requires a tunable Hamiltonian by switching to another set of laser frequencies on the cavity, which is not always accessible.  A more general investigation of the LMG model in spin squeezing is required and whether it could reach Heisenberg-limited noise reduction remains unknown.

Here we study the spin squeezing properties in the LMG model with different
anisotropic parameters. We find that initial state and the anisotropic
parameter play important roles in the spin squeezing. We propose an
implementable way to transform the LMG model into effective TAT model by
making use of rotation pulses along different axes on the Bloch sphere,
which gives a convenient way to generate efficient spin squeezing reaching
the HL. We also analyze the influence of noises and find that our scheme is
robust to fluctuations in pulse areas and pulse separations.


%
%
%

The paper is organized as follows. In Sec.~(\ref{sec:1}), we first introduce
the system model of the quadratic collective-spin interaction, which can be
described by the LMG model.
In Sec.~(\ref{sec:2}), we investigate the performance of spin squeezing in
the LMG mode and present the optimal initial state for spin squeezing in the
LMG model. In Sec.~(\ref{sec:3}), we prove that the designed rotation pulse
method can transform the LMG model into effective TAT interaction. We also
show that the method is robust to different noises according to numerical
simulations.

\section{\label{sec:1}THE SYSTEM MODEL}


We consider a system of mutually interacting spin-1/2 particles described by
the following Hamiltonian:
\begin{equation}
H = \sum_{j<k}\chi_{\alpha\beta}\sigma_{\alpha}^{j}\sigma_{\beta}^{k},
\label{eq:one}
\end{equation}
where $\sigma_{\alpha}^{j}$ is the Pauli operator of the $j$-th spin and $%
\alpha,\beta\in\{x,y,z\}$. The parameter $\chi_{\alpha\beta}$ characterize
the strength of the interaction in different directions. To ensure the
Hermicity of the Hamiltonian, we have $\chi_{\alpha\beta} =
\chi_{\beta\alpha}$. Here we have the assumption that the interactions between individual spins are the same. This assumption holds when there are all-to-all interactions rather than just dipole-dipole interactions, which is valid under some systems such as nuclear system \cite{27}, Cavity QED \cite{9}, ion trap \cite{10}.
Now we introduce the collective spin operators $S_{\alpha} = \frac{\hbar}{2}
\sum_{j}\sigma_{\alpha}^{j}$. Let $\hbar = 1$, using
\begin{equation}
\sigma_{\alpha}\sigma_{\beta} =
i\sum_{\gamma}\varepsilon_{\alpha\beta\gamma}\sigma_{\gamma}+\delta_{\alpha%
\beta},
\end{equation}
where $\varepsilon_{\alpha\beta\gamma}$ is the Levi-Civita symbol and $%
\delta_{\alpha\beta}$ is the Kronecker delta, Eq.~(\ref{eq:one}) becomes
\begin{equation}
H =
2\sum_{\alpha,\beta\in\{x,y,z\}}\chi_{\alpha\beta}S_{\alpha}S_{\beta}+H_{0},
\end{equation}
where $H_{0}$ is a constant and can be neglected. $H$ preserves the
magnitude of the total spin $S^{2} = \sum_{\alpha}S^{2}_{\alpha} $, namely,%
%
%
\begin{equation}
[H,S^2] = 0,
\end{equation}
which means $S^{2}$ is a constant. The Hamiltonian can be written as
\begin{equation}
H = \boldsymbol{S^{T}A S},
\end{equation}
where $A_{\alpha\beta} = 2\chi_{\alpha\beta}$. $\boldsymbol{A}$ is a real
symmetric matrix, which means it can be diagonalized by a linear
transformation:
\begin{equation}
\boldsymbol{A} = \boldsymbol{Q^{-1}D Q},
\end{equation}
in which $\boldsymbol{Q}$ is an orthogonal matrix and $\boldsymbol{D}$ is a
diagonal matrix whose nonzero elements are eigenvalues of $\boldsymbol{A}$.
Let $\boldsymbol{\tilde{S}} = \boldsymbol{QS}$, we can turn the Hamiltonian
into the canonical form: 
\begin{equation}
H = \sum_{\alpha\in\{x,y,z\}}\chi_{\alpha}\tilde{S}_{\alpha}^{2},
\end{equation}

For convenience, in the following we redefine the spin operator $S_{\alpha}$
by omitting the tilde.
We select the corresponding $\tilde{S_{\alpha}}$ of the largest $%
\chi_{\alpha}$ as $S_{x}$ and the minimum as $S_{z}$, i.e., $%
\chi_{x}\geq\chi_{y}\geq\chi_{z}$. Using the relation $\sum_{\alpha}S^{2}_{%
\alpha} = S^{2}$, the transformed Hamiltonian reads
\begin{equation}
H = (\chi_{x}-\chi_{z})S_{x}^{2}+(\chi_{y}-\chi_{z})S_{y}^{2}+S^2.
\end{equation}
Let $\chi = \chi_{x}-\chi_{z}$, $\gamma =
(\chi_{y}-\chi_{z})/(\chi_{x}-\chi_{z})$ and ignore the constant term, we
obtain the general form of the Hamiltonian of the LMG model
\begin{equation}
H = \chi(S_{x}^{2}+\gamma S_{y}^{2}) .  \label{eq:9}
\end{equation}
Therefore, any system with Hamiltonian in the form of Eq.~(\ref{eq:one}) can be
transformed to the standard form of the LMG model as Eq.~(\ref{eq:9}). What's worth mentioning is that we ignore the linear interaction between the spin and external magnetic field. The reason is that linear interaction itself can't generate spin squeezing, and the linear interaction could be easily canceled in the experimental system using suitable pulse sequences.

Under the condition $\chi_{x}\geq\chi_{y}\geq\chi_{z}$, we have $0\le\gamma
\le1$. Furthermore, note that if $0.5<\gamma \le 1$, we have:
\begin{equation}
H = \chi(S_{x}^{2}+\gamma S_{y}^{2} - S^2) = -\chi[S_{z}^2 +
(1-\gamma)S_{y}^2].
\end{equation}
which is equivalent to $\chi [S_{x}^2 + (1-\gamma)S_{y}^2]$ if we switch the
$x$-axis and the $z$-axis. Hence, we only need to consider the situation
when $0\le \gamma \le 0.5$. Specially, when $\gamma=0$ ($\gamma=0.5$), the
LMG Hamiltonian reduces to the OAT (TAT) Hamiltonian.

\section{\label{sec:2}SPIN SQUEEZING OF THE LMG MODEL}

To describe the properties of SSS, we investigate the squeezing parameter
given by Kitagawa and Ueda \cite{2}:
\begin{equation}
\xi^2 = \frac{4(\Delta S_{\vec n_{\perp}})^2}{N},
\end{equation}
where subscript $\vec n_{\perp}$ refers to an arbitrary axis perpendicular
to the mean spin direction, where the minimum value of $(\Delta S)^2$ is
obtained. The inequality $\xi^2 < 1$ indicates that the state is squeezed.

\begin{figure}[tb]
\centering
\includegraphics[width = 0.5\textwidth]{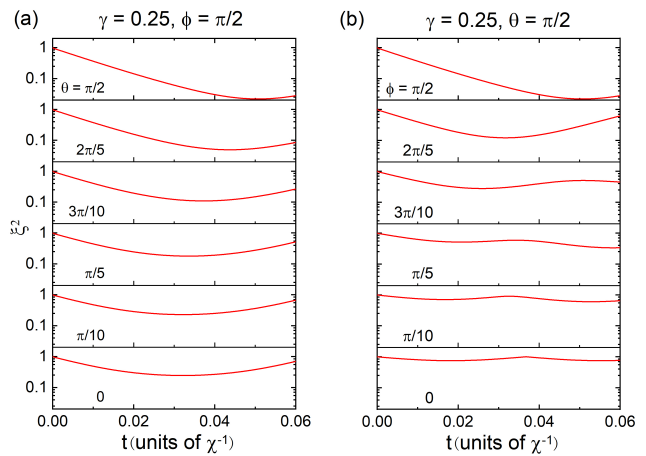}
\caption{Time evolution of the squeezing parameter $\protect\xi^2$ for
anisotropic parameter $\protect\gamma=0.25$ with different initial $\protect%
\phi$ and $\protect\theta$. (a) $\protect\phi = \protect\pi/2$ , $\protect%
\theta$ changes from 0 to $\protect\pi/2$. (b) $\protect\theta = \protect\pi%
/2$ , $\protect\phi$ changes from 0 to $\protect\pi/2$. }
\label{fig:11}
\end{figure}

\begin{figure}[b]
\includegraphics[width = 0.5\textwidth]{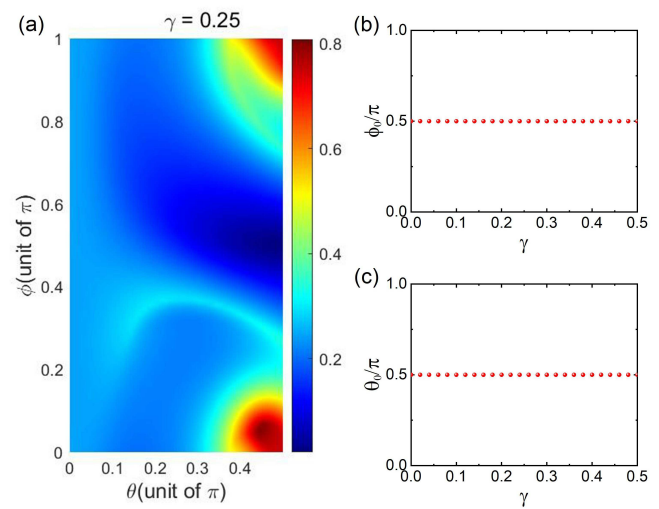}
\caption{Minimum squeezing parameter $\protect\xi^2$ as a function of the
initial $\protect\theta$ and $\protect\phi$. (b)The optimal initial state
azimuth angle $\protect\phi_0$ when $\protect\xi^2$ reaches its minimum in
the condition of different $\protect\gamma$. (c)The optimal initial state
polar angle $\protect\theta_0$ when $\protect\xi^2$ reaches its minimum in
the condition of different $\protect\gamma$. The anisotropic parameter is $%
\protect\gamma=0.25$.}
\label{fig:12}
\end{figure}

The Hamiltonian of the LMG model is a typical kind of nonlinear interaction,
which produces SSS by time evolution. We choose the coherent spin states as
the initial states, which can be described by
\begin{equation}
\ket{\theta,\phi} = e^{i\theta(S_{x}\sin\phi - S_{y}\cos\phi)}\ket{j,j},
\end{equation}
where $\theta$ is the angle between the $z$-axis and the collective-spin
vector (polar angle), while $\phi$ is the angle between the $x$-axis and the
vertical plane containing the collective-spin vector (azimuth angle).

Typical examples of the time evolution of $\xi^2$ are presented in Fig.~\ref%
{fig:11}. It reveals that the squeezing parameter reaches a local minimum in
a short time scale. For a certain $\gamma$, the minimum squeezing parameter
and the corresponding time varies with the initial $\theta$ and $\phi$.

In Fig.~\ref{fig:12}(a), we plot the color map of the minimum squeezing
parameter as functions of the initial $\phi$ and $\theta$ for fixed $\gamma$
(for example, $\gamma = 0.25$). It reveals that the optimal initial state
with best squeezing is $\ket{\theta,\phi}=\ket{\pi/2,\pi/2}$. Similarly, we
change $\gamma$ and plot the color maps, with the optimal initial $%
\phi $ and $\theta$ plotted in Fig.~\ref{fig:12}(b) and Fig.~\ref{fig:12}%
(c).
We can see that when $\gamma$ varies from 0 to 0.5, the LMG model obtains
the optimal squeezing when the initial state is $\ket{\pi/2,\pi/2}$. This
can be understood in an intuitive sense: when $0\le \gamma \le 0.5 $, we
have:
\begin{equation}
H = \chi[(1-\gamma)S_{x}^{2} - \gamma S_{z}^{2}+\gamma S^2],
\end{equation}
which can be seen as two counter-twisting squeezing acting around the $x$%
-axis and $z$-axis, respectively, and along the $y$-axis these two effects
reach the optimal cases at the same time. Thus the optimal initial state is
always $\ket{\pi/2,\pi/2}$.

\begin{figure}[tb]
\includegraphics[width = 0.5\textwidth]{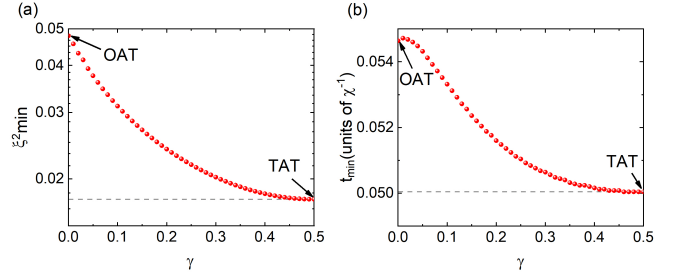}
\caption{(a) Minimum squeezing parameter for different anisotropic parameter
$\protect\gamma$. (b) The corresponding time it takes to get the minimum
squeezing parameter for different $\protect\gamma$. The horizontal dashed
lines correspond to the results of TAT model.}
\label{fig:13}
\end{figure}

Now we let the initial state be the optimal case $\ket{\pi/2,\pi/2}$, which could be realized through optical pumping and a $\pi/2$ pulse along the $x$-axis. And we track how the minimum squeezing parameter $\xi^2$ changes when $\gamma$
varies from 0 to 0.5. The results are shown in Fig.~\ref{fig:13}. We can
conclude that the squeezing performance monotonically depends on $\gamma$
for $0 \leq \gamma \leq 0.5$. When $\gamma = 0.5$, the LMG model attains its
minimum $\xi^2$, and the corresponding time is also the shortest,
corresponding to the TAT squeezing.
Therefore, to reach the best squeezing performance, we should ensure the
anisotropic parameter approaches $\gamma=0.5$.

However, when the anisotropic parameter takes other values, the squeezing
performance degrades. To solve this problem, we propose to introduce
rotation pulses capable of transforming the LMG model into TAT model.


\section{\label{sec:3}Transforming LMG INTO TAT}

Inspired by the previous study of transforming the OAT interaction into the
TAT interaction \cite{24}, our idea is to transform the LMG model into the
TAT model by making use of multiple ${\pi}$/2 pulses, which can be realized
using the coupling term $\Omega_{\alpha}S_{\alpha}$ ($\alpha=x,y,z$). In the
Rabi limit $\left| \Omega \right| \gg N \left| \chi \right| $, nonlinear
interaction can be neglected while the collective spin undergoes driven Rabi
oscillation. By making use of a multi-pulse sequence along the $\alpha$-axis
($\alpha=x,y,z$), we can rotate the spin along the $\alpha$-axis and affect
the dynamic of squeezing. A $\pi/2$ pulse corresponds to $%
\int_{-\infty}^{+\infty} \Omega_{\alpha}(t)dt=\pi/2 $, which leads to the
result that $R_{\alpha,-\pi/2}e^{iS^{2}_{\beta}}R_{\alpha,\pi/2}=e^{iS^{2}_{%
\kappa}}$, where $R_{\alpha,\theta}=e^{-i\theta S_{\alpha}}$ and $\kappa$ is
the axis that perpendicular to the $\alpha$-axis and $\beta$-axis. The
multi-pulse sequence is periodic, and the frequency is determined by $\gamma$
and the axis we choose.

\begin{figure}[b]
\includegraphics[width = 0.45\textwidth]{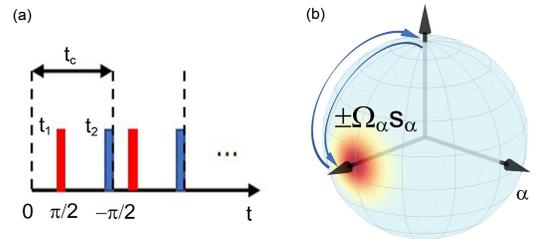}
\caption{(a) An illustration of the pulse sequences. The overall process
could be viewed as the repetition of (a). (b) The pulse method could also be
viewed as the cyclic rotation around the $\protect\alpha$-axis on the Bloch
sphere. }
\label{fig:14}
\end{figure}

As shown in Fig.~\ref{fig:14} (a), each period is made up of the following:
a $-\pi/2$ pulse along the $\alpha$-axis, a free evolution for $\delta t_1$,
a $\pi/2$ along the $\alpha$-axis, and a free evolution for $\delta t_2$. The
period is $t_c=\delta{t_1}+\delta{t_2}$, neglecting the time needed for
applying the two $\pi/2$ pulses. Figure~\ref{fig:14} (b) shows that the
cyclic $\pm \pi/2$ could be viewed as rotations on the Bloch sphere. By
adjusting the relationship between $\delta t_1$ and $\delta t_2$, we can
transform the LMG model into TAT. One general Hamiltonian for TAT interaction is $H_{\mathrm{TAT}}=S_{x}S_{y}+S_{y}S_{x}$ for $\theta=\pi/2, \phi=\pm \pi/4$ \cite{2}. By changing the initial states and twisting axes, the TAT interaction could also be expressed as ${\ H_{%
\mathrm{TAT}}} \propto {S_y^{2}-S_x^2}$, ${H_{\mathrm{TAT}}} \propto {%
S_z^{2}-S_y^2}$ and ${H_{\mathrm{TAT}}} \propto {S_x^{2}-S_z^2}$. In Bloch
sphere, the first expression indicates $\phi=\pi/2,\theta=0$, while the
middle expression indicates $\phi=0,\theta=\pi/2$ and the last expression
indicates that $\phi=\pi/2,\theta=\pi/2$. According to $S^{2}_{x}+S^{2}_{y}+S^{2}_{z}=S^2$, $S^{2}_{x}-S^2_{z}=2(S^2_{x}+0.5S^2_{y})-S^2$, $S^2$ will not influence the properties of spin squeezing, we simply ignore it.

\begin{figure}[tb]
\includegraphics[width=0.5\textwidth]{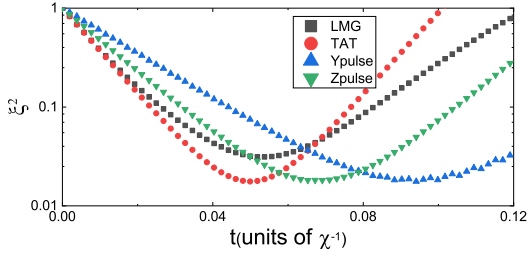}
\caption{Numerical analysis of squeezing parameter of the LMG model (Black
Square), TAT (Red Circle), the LMG model after pulse sequences along $y$%
-axis (Blue Triangle-Up), and the LMG model after pulse sequences along $z$%
-axis (Green Triangle-Down) with $N=100, \protect\gamma=0.1$. For $0\le
\protect\gamma \le 0.5$, pulse sequences along the $z$-axis will have higher
squeezing strength, thus leading to faster squeezing.}
\label{fig:15}
\end{figure}

\begin{figure}[b]
\includegraphics[width=0.5\textwidth]{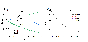}
\caption{(a) Squeezing time of the $y$-pulse method (Blue Triangle-Up), and $%
z$-pulse method (Green Triangle-Down) with different $\protect\gamma$.
Insets: Effective squeezing strength of the $y$-pulse and $z$-pulse methods.
(b) Squeezing ratio of the OAT, TAT, LMG, and pulse method. }
\label{fig:16}
\end{figure}

\begin{figure}[tb]
\centering
\includegraphics[width=0.5\textwidth]
{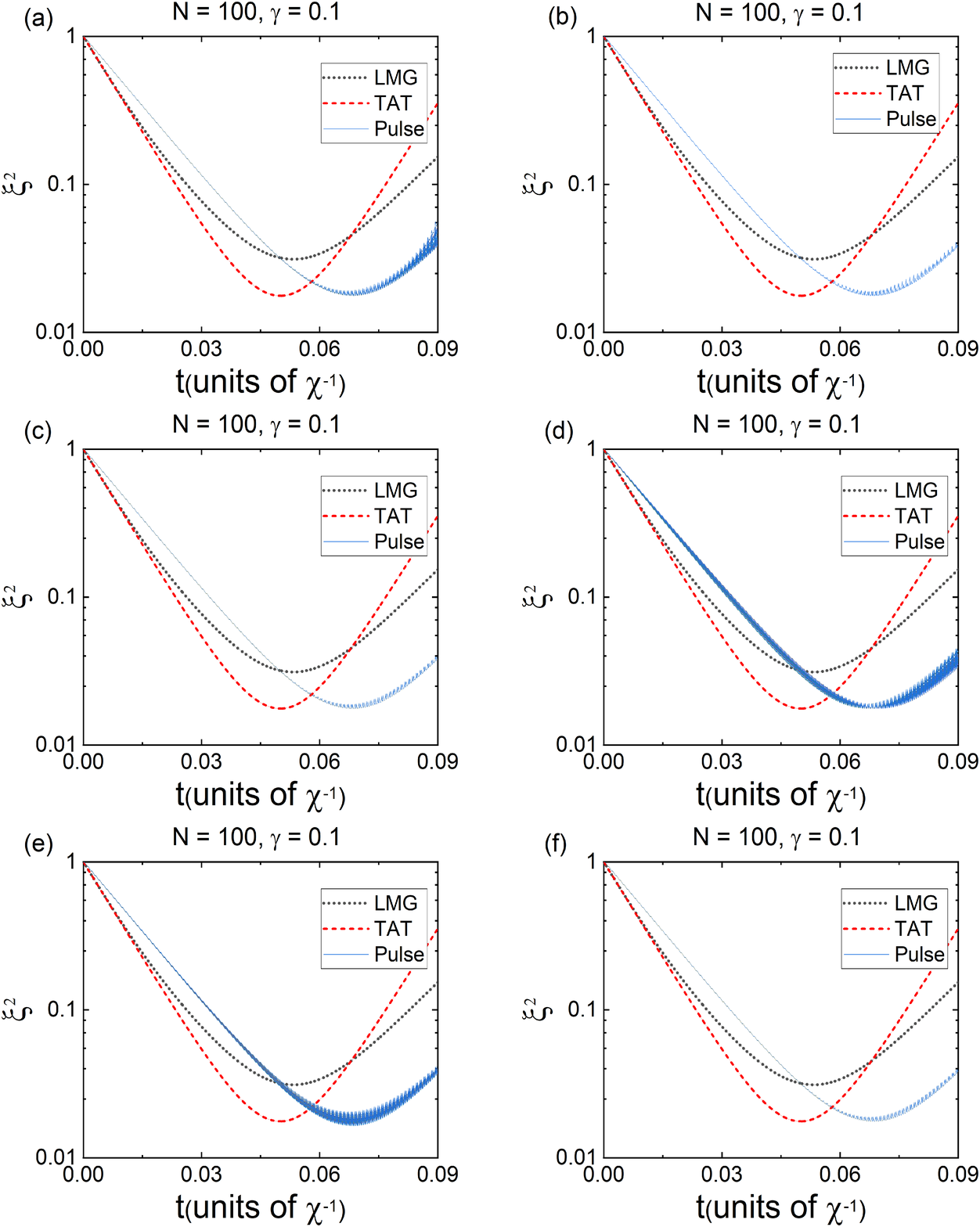}
\caption{Numerical analysis of the influence of noises for our scheme with $%
N=100,\protect\gamma=0.1$. The blue curves crowding together denote the
results of 100 independent simulations under different noises. (a) Evolution
of the squeezing parameter $\protect\xi^2$ for 10\% level of Gaussian
stochastic noise adding on the pulse separation. (b) Evolution of the
squeezing parameter $\protect\xi^2$ for 0.02\% level of Gaussian stochastic
noise adding on the pulse area. (c) Evolution of the
squeezing parameter $\protect\xi^2$ for 0.01\% level of Gaussian stochastic
noise adding on $\gamma$. (d) Evolution of the
squeezing parameter $\protect\xi^2$ for 1\% level of Gaussian stochastic
noise adding on interaction strength $\chi$. (e) Evolution of the
squeezing parameter $\protect\xi^2$ for 0.01\% level of Gaussian stochastic
noise adding on atom number $N$. (f) Evolution of the
squeezing parameter $\protect\xi^2$ for 0.1\% level of Gaussian stochastic
noise adding on pulse stability.}
\label{fig:17}
\end{figure}
For ${H_{\mathrm{LMG}}}=\chi({S_x^{2}+{\gamma}S_y^2})$, if we choose the $z$%
-axis to be the $\alpha$-axis, the time evolution operates for a single
period is the following:
\begin{eqnarray}
{U_z}&=&e^{-i(S^2_x+\gamma S^2_y)\chi t_1}R_{z,-\pi/2}e^{-i(S^2_x+\gamma
S^2_y)\chi t_2}R_{z,\pi/2}  \notag \\
&=&e^{-i(S^2_{x}+ \gamma S^2_{y})\chi t_1}e^{-i(S^2_{y}+ \gamma S^2_{x})\chi
t_2.}
\end{eqnarray}

Using the Baker-Campbell-Hausdorff formula, we find $U_{z}\approx e^{-i\chi
(S_{x}^{2}(t_{1}+\gamma t_{2})+S_{y}^{2}(\gamma t_{1}+t_{2}))}$ for small t.
To transform the LMG model into TAT twisting, the coefficients should
satisfy
\begin{equation}
\frac{t_{1}+\gamma t_{2}}{\gamma t_{1}+t_{2}}=0.5\text{ or }2.
\end{equation}%
Then the relationship between $\gamma $ and $t_{2}/t_{1}$ should satisfy
\begin{equation}
\frac{t_{2}}{t_{1}}=\frac{\gamma -2}{2\gamma -1}\text{ or }\frac{2\gamma -1}{%
\gamma -2}.
\end{equation}%
Accordingly, we obtain the effective Hamiltonian%
\begin{eqnarray}
H_{z}^{\mathrm{eff}} &=&\frac{\chi (\gamma +1)}{3}(S_{x}^{2}+2S_{y}^{2}), \\
\text{or }H_{z}^{\mathrm{eff}} &=&\frac{\chi (\gamma +1)}{3}%
(2S_{x}^{2}+S_{y}^{2}).
\end{eqnarray}

Similarly, if we choose the $y$-axis to be the $\alpha $-axis, we have the
time evolution operator for a single period:
\begin{eqnarray}
{U_{y}} &=&e^{-i(S_{x}^{2}+\gamma S_{y}^{2})\chi t_{1}}R_{y,-\pi
/2}e^{-i(S_{x}^{2}+\gamma S_{y}^{2})\chi t_{2}}R_{y,\pi /2}  \notag \\
&=&e^{-i(S_{x}^{2}+\gamma S_{y}^{2})\chi t_{1}}e^{-i(S_{z}^{2}+\gamma
S_{y}^{2})\chi t_{2}}
\end{eqnarray}%
To achieve TAT, we require $t_{2}/t_{1}=(\gamma +1)/(-\gamma +2)$ or $%
t_{1}/t_{2}=(\gamma +1)/(-\gamma +2)$. Then the resultant Hamiltonians read

\begin{eqnarray}
H_{y}^{\mathrm{eff}} &=&\frac{\chi (1-2\gamma )}{3}(2S_{x}^{2}+S_{y}^{2}), \\
\text{or }H_{y}^{\mathrm{eff}} &=&\frac{\chi (2\gamma -1)}{3}%
(S_{x}^{2}+2S_{y}^{2}).
\end{eqnarray}%
However, if we choose the $x$-axis to be the $\alpha $-axis, we will find
that for $0\leq \gamma \leq 0.5$, $t_{1}/t_{2}\leq 0$, which means it's
impossible to transform the LMG model into TAT twisting making use of
multi-pulse sequence along $x$-axis. The above pulse sequences are
numerically verified with the results present in Fig.~\ref{fig:15}.

To make the squeezing occur faster, we need to shorten the squeezing time,
which means getting higher squeezing strength $\chi^{\mathrm{eff}}$. As Fig.~%
\ref{fig:16} (a) shows, for $0\le \gamma \le 0.5$, pulse along $z$-axis gets
higher effective strength, which is $\chi^{\mathrm{eff}}_{z}=\chi(1+%
\gamma)/3 $. And the squeezing time of the $z$-pulse method is also shorter
compared with the $y$-axis pulse method.

Therefore, for a LMG model with arbitrary anisotropic parameter $\gamma$
ranging from 0 to 0.5, we can transform it into TAT interaction by using
multi-pulse along different axes, and the squeezing performance of the LMG
model after the pulse sequences will also reach Heisenberg scaling, as good
as the TAT case, as compared in Fig.~\ref{fig:16} (b).

Our scheme is robust to different kinds of noises. We carry out the numerical simulation by adding
Gaussian stochastic noises, i.e., assuming the fluctuating pulse areas, pulse separations, pulse stability, $\gamma$, atoms number $N$, and interaction strength $\chi$,   are subject to Gaussian distribution with a standard
deviation of different ranges of the average value. The squeezing parameters of
100 independent simulations under different types of noises are respectively
shown in Fig.~\ref{fig:17}. The numerical simulations show that our method is not only robust to internal systems noise such as uncertainty of determining $\gamma$, uncertainty of determining atoms number $N$,  uncertainty of interaction strength $\chi$ but also external noise such as pulse areas noise, pulse separation noise, and pulse phase instability.     Under certain kinds and ranges of noise, the best attainable squeezing of our method can almost achieve the optimal
squeezing of the effective TAT dynamics.

As for the spin decoherence, our method itself will not bring new resources of decoherence but extend the squeezing time, while the coherence time for atoms such as Dysprosium is very long in spin squeezing \cite{34}, so we ignore the impact of extending evolution time for spin decoherence.

\section{Conclusion}

In conclusion, we study the spin squeezing in systems with quadratic
collective-spin interaction, which can be described by the LMG model. We
find that the squeezing performance depends on the initial state and the
anisotropic parameter. We show that the best initial state for $H=\chi
(S^2_x+\gamma S^2_y)$ is $\ket{\pi/2,\pi/2}$, which holds for different
anisotropic parameter $\gamma$.
We propose an implementable way with rotation pulses to transform the LMG
model into TAT model with Heisenberg-limited spin squeezing. We find that
pulse sequences applied along the $z$-axis will result in larger squeezing
strength $\chi^{\mathrm{eff}}_{z}=\chi(1+\gamma)/3$ compared to the pulse
sequences along the $y$-axis $\chi^{\mathrm{eff}}_{y}=\chi (1-2\gamma)/3$.
Besides, our scheme is robust to noise in pulse areas and pulse separations.
Our work will significantly increase the systems that could reach the
Heisenberg scaling and will push the frontier of quantum metrology.


\section{acknowledgement}

We thank Prof. Fu-li Li for helpful discussions. This work is supported by
the Key-Area Research and Development Program of Guangdong Province (Grant
No. 2019B030330001), the National Natural Science Foundation of China (NSFC)
(Grant Nos. 12275145, 92050110, 91736106, 11674390, and 91836302), and the
National Key R\&D Program of China (Grants No. 2018YFA0306504).

\end{document}